\documentclass[10pt]{iopart}
\renewcommand{\thesection}{\arabic{section}}
\renewcommand{\thesubsection}{\arabic{section}.\arabic{subsection}}
\usepackage{mathrsfs}
\usepackage{amssymb}
\usepackage{dsfont}
\usepackage{graphicx}
\usepackage[usenames,dvipsnames,svgnames,x11names,table]{xcolor}
\usepackage{bm}
\usepackage{subfigure}
\usepackage{float}
\restylefloat{table}
\usepackage{ulem}

\begin{document}

\title{On the oscillating properties of a two-electron quantum dot  in the presence of a magnetic field}
 \author{A. M. Maniero$^1$, C. R. de Carvalho$^2$, F. V. Prudente$^3$ and Ginette Jalbert$^2$}
 \address{ 
$[1]$ Centro das Ciências Exatas e das Tecnologias, Universidade Federal do Oeste da Bahia, 47808-021,
Barreiras, BA, Brazil\\
$[2]$ Instituto de F\'{\i}sica, Universidade Federal do Rio de Janeiro, Rio de Janeiro, 21941-972, RJ, Brazil\\
$[3]$ Instituto de F\'{\i}sica, Universidade Federal da Bahia, Campus Universit\'ario de Ondina, 40170-115, Salvador, BA, Brazil}
\ead{ginette@if.ufrj.br}

\begin{abstract}
 We give a basic explanation for the oscillating properties of some physical quantities of a two-electron quantum dot in the presence of a static magnetic field. This behaviour was discussed in a previous work of ours [AM Maniero, {\it et al}. J. Phys. B: At. Mol. Opt. Phys. 53:185001, 2020]  and was identified as a manifestation of the {\it de Haas-van Alphen} effect, originally observed in the framework of diamagnetism of metals in the 30's. We show that this behaviour is a consequence of different eigenstates of the system assuming, in a certain interval of the magnetic field, the condition of the lowest energy singlet and triplet states.
\end{abstract}

%\submitto{\jpb}
 \maketitle
 
\ioptwocol

Recently in a series of articles we have studied the behaviour of electrons in quantum dots (QDs) with different confinement profiles and under the influence of external fields \cite{olavo2016, maniero2019, maniero2020a}. By addressing the system constituted of two electrons confined in a QD in the presence of a static external magnetic field we have been faced with a peculiar behaviour of some physical quantities associated with the two electrons that has been known for a long time in the scope of metal diamagnetism. We have observed that some characteristic quantities of the system -- such as the exchange energy $J=E_T - E_S$ ($E_S$ and $E_T$ are the energies of the lowest singlet and triplet states, respectively), and the density $\rho(x_1,x_2)$ of the electronic cloud along one of the axes perpendicular to the magnetic field ($\vec B=B\hat z$) -- display an oscillating behaviour as function of the field magnitude. This behaviour was originally observed experimentally by de Haas and van Alphen  when they were studying the dependence with the magnetic field of the susceptibility of diamagnetic metals \cite{deHaas1930}. By that time the diamagnetism of metals had already been studied theoretically by several researchers and the model adopted was a free electron gas \cite{Fock1928, Landau1930, Darwin1931}. The phenomenon observed by de Haas and van Alphen aroused even more interest in the quantum description of a free electron system and numerous researchers have addressed the problem, see for instance Refs. \cite{Sondheimer1951, Dingle1952} and references therein. More recently, this oscillating behaviour associated with the properties of an electron system was theoretically predicted in the context of quantum dots in semiconductors. The Coulomb interaction between electrons, which until then has been disregarded according to the free electron model, is taken into account \cite{maksym1990, wagner1992}.

Therefore, the Hamiltonian of a two-electron anisotropic 3D harmonic QD is given by 
\begin{eqnarray}
	\hat H =\sum_{j=1}^2 & \frac{1}{2m_c}\big[\vec p_j + \vec A(\vec r_j) \big]^2 +  g\mu_B\vec S_j \cdot \vec B_j  + \hat{V}(\vec r_j) \nonumber \\ 
	& +\frac{1}{\kappa \vert \vec r_1-\vec r_2 \vert},
\end{eqnarray}
where 
\begin{eqnarray}
\hat{V}(x,y,z) = \frac{m_c}{2}\left(\omega^2_x x^2+\omega^2_y y^2+\omega^2_z z^2\right)
\label{Vq}
\end{eqnarray}
is the anisotropic 3D harmonic confining potential and the vector potential is choosen in the gauge $\vec\nabla \cdot \vec A = 0$, which for $\vec B = B\hat{z}$ yields
\begin{eqnarray}
\vec A(\vec r)= -\frac{1}{2}\vec r \times \vec B = \frac{1}{2}\big( - y\hat{x} + x\hat{y} \big)B.
\label{vecA}
\end{eqnarray}
One also has that
\begin{eqnarray}
 \frac{1}{2m_c}\big(\vec p_j + e\vec A(\vec r_j) \big)^2 =& -\frac{\nabla_j^2}{2m_c} + \frac{\vec B \cdot \vec L}{2m_c} \nonumber \\ 
&+  \frac{ B^2}{8m_c}(x^2 + y^2),
\label{(p2+eA)^2/2m}
\end{eqnarray}
which allows one to write the Hamiltonian in the form
\begin{eqnarray}
\hspace{0.8cm} \hat H &=&\sum_{j=1}^2\Big[-\frac{\nabla_j^2}{2  m_c}  +\frac{m_c}{ 2 } \Big(\Omega_{xL}^{2} x_j^2  +\Omega_{yL}^{2} y_j^2
\nonumber \\
&& +\omega_{z}^{2} z_j^2 \Big) + \frac{\mu_BB}{m_c}  \hat L_{jz} \Big] + \mu_BB  g \hat S_{z}
\nonumber \\
&&  +\frac{1}{\kappa \vert \vec r_1-\vec r_2 \vert},
\label{hamiltonian}
\end{eqnarray}
where $ \Omega^2_{xL} = (\omega^2_x +\omega_L^2)$ and  $\omega_L=B/2 m_c$ is  the Larmor frequency. For more details of the present theoretical approach, see Ref.~\cite {maniero2020a}. There, we compute the solution $\Phi$ of the  Sch\"odinger equation within the framework of the full configuration interaction method (full-CI), employing the {\it Cartesian anisotropic Gaussian-type orbitals} as basis functions. In this method $\Phi$ is written as a linear combination of configuration state functions (CSFs) which, in turn, are constituted of a linear combination of Slater determinants. In this previous work we compared the precision or the reliability of the results obtained from three different basis sets: 2s2p2d, 2s2p2d2f, and 2s2p2d2f1g. The last one, consisting of 55 basis functions, leads to 1485 (1485)  CSFs and 2916 (1485) determinants for the singlet (triplet) states.  In the present article, we use only this latter one, the largest basis set, for all numerical computation.

From the solution $\Phi$ we can compute some physical quantities of interest such as the root-mean-square $\sigma_x = \sqrt{\langle x^2 \rangle}$, which gives information on the spatial spreading of the two electrons along a direction perpendicular to the magnetic field, as well as the $z$-component of the total orbital angular momentum of the two electrons, $L_z$. In our previous paper \cite{maniero2020a} we  performed an approximation in the expansion of $\Phi$ by taking into account only the CSFs whose coefficients had modulus greater than an arbitrary value. To accomplish the present work we introduced modifications in the numerical code that allowed us to take into account all the 1485 CSFs, and still with a smaller computation time. These improvements in our code should be the subject of a future paper.

With these numerical modifications we verified that the previously obtained  values of the system physical quantities as a function of the magnetic field $B$ did not change and the gain in the computation time allowed us to study in much more detail their behaviour around the extremes of $J$, i.e. $dJ/dB=0$ (in fact we have dealt with $J_{norm}=J/\omega_x$) . It is worth remembering that we identified a correlation between the $J_{norm}$ extremes and the crossing of the $\sigma_x$ curves for the singlet and triplet states~\cite{maniero2020a}. In the present Letter we explore this issue.

\begin{figure}[t]
%\vspace{-0,5 cm} height=8 cm,width=\columnwidth
%\includegraphics[scale=0.25]{Fig1_3D-JVL.pdf}
\includegraphics[height=8 cm,width=\columnwidth]{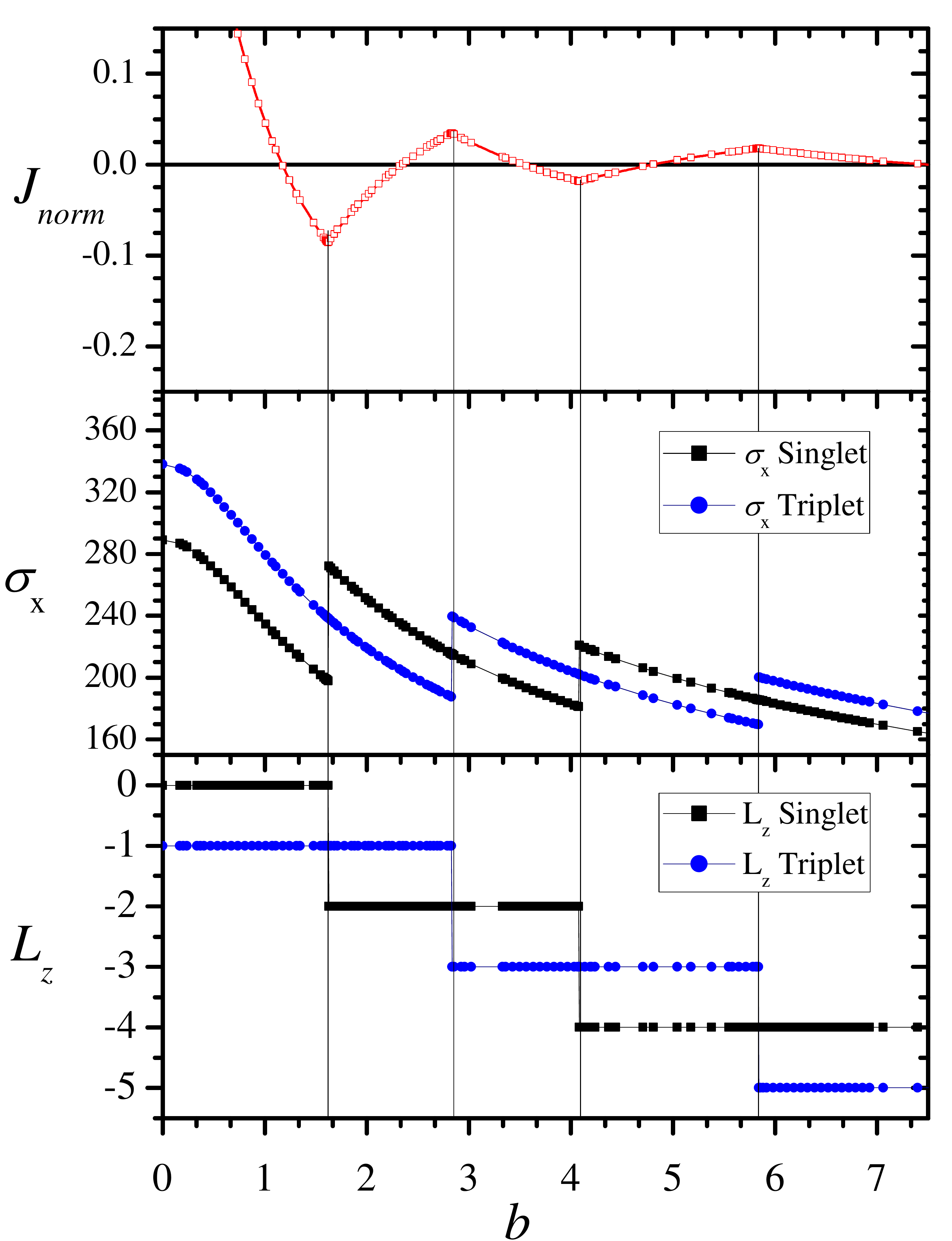}
\vspace{-0.5 cm}
\caption {It is displayed $J_{norm}$ , $\sigma_x$ and $L_z$ of the singlet and triplet states of lowest energy, as a function of $b$, for the confinement condition given by $\omega_x=\omega_y=\omega_z=0.000111$ (3D case). As highlighted in the legend the solid squares correspond to the singlet state, whereas the solid squares to the triplet state.}
\label{3DJLz}
\end{figure}

All the results displayed in this article were obtained with the same values employed in our previous work , in particular, $\omega_x$ = $\omega_y$ = $\omega_z$ = 0.000111 (3D case), and $\omega_x$ = $\omega_y$ = 0.000111 with $\omega_z$ = 1.11 ($quasi$-2D case). The computations have been done in atomic units (au), as usual in atomic-molecular calculations. The value $\omega_x= 0.000111$ corresponds to an energy ($\hbar \omega_x$) of 3 meV~\cite{carvalho2003}. The magnetic field is given in units of a characteristic magnetic field $B_c  = 2 m_c \omega_x$, where $m_c$ is the effective electronic mass which yields 0.067 for the conduction band in GaAs and leads to $B_c \approx 1.49\times 10^{-5}$ ( $\approx 3.5$ T). 

\begin{figure}[t]
%\vspace{-0,5 cm}
%\includegraphics[scale=0.35]{Fig2_2D-JVL.pdf}
\includegraphics[height=8 cm,width=\columnwidth]{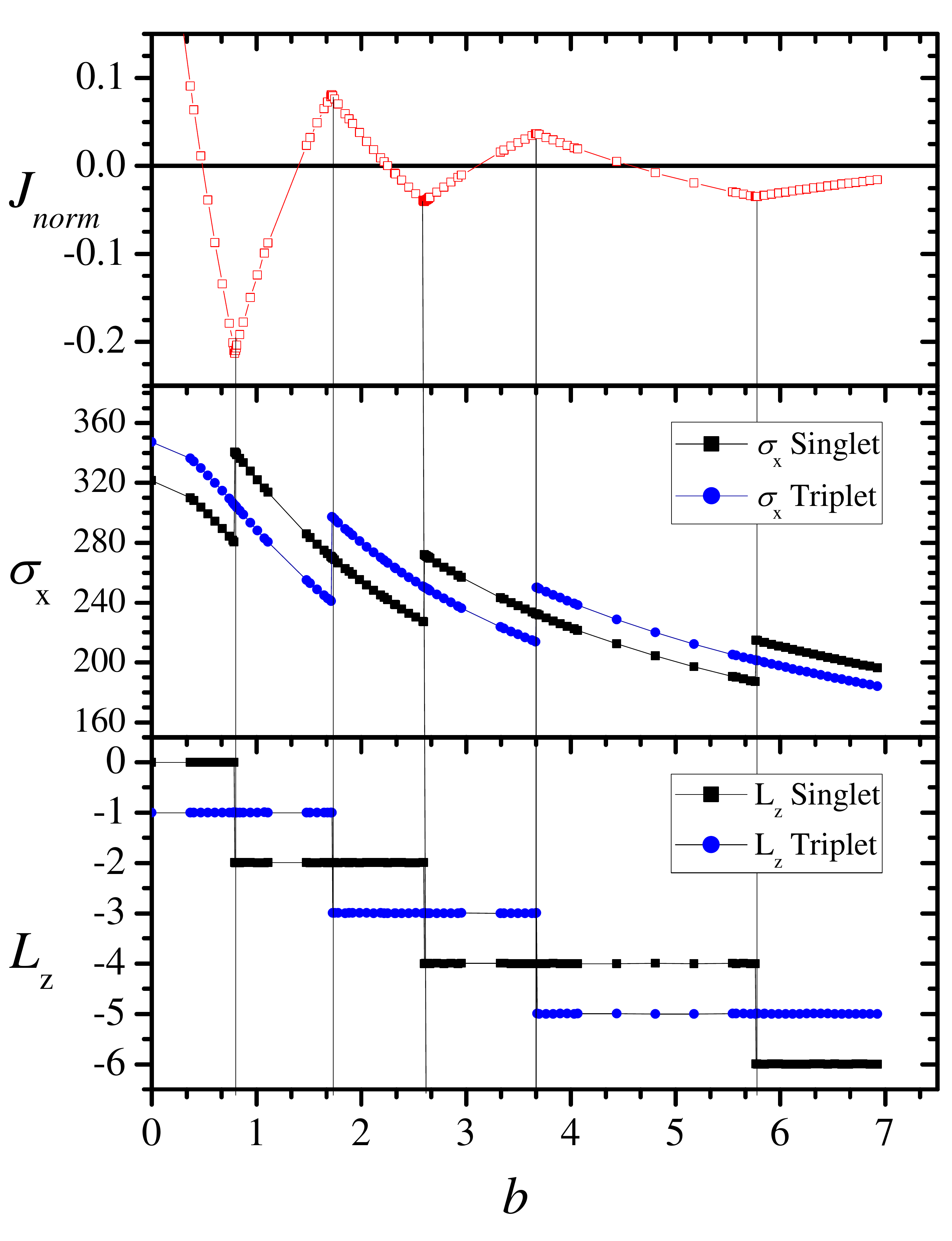}
\vspace{-0.5 cm}
\caption {It is displayed the same as in the previous figure, except that the confinement condition correspond to the $quasi$-2D case, i.e., $\omega_x=\omega_y=0.000111$ and  $\omega_z=1.11$.}
\label{2DJLz}
\end{figure}

 Here, at this point, it is worth noting that the operators $\hat H$ and $\hat L_z$ commute and, therefore, the eigenstates of $\hat H$ are also eigenstates of $\hat L_z$. In  Figs.~\ref{3DJLz} and~\ref{2DJLz} we display the behaviour of  $J_{norm}$, $\sigma_x$ and $L_z$ of the singlet and triplet states of lowest energy as function of the normalized field $b$ ($b=B/B_c$) for $\omega_z=0.000111$ and $\omega_z=1.11$, respectively. One can see the connection between the $J_{norm}$ extremes and the changes of the $\sigma_x$ values of the singlet and triplet states, which in turn reflects the changing in their $z$-component of the orbital angular momentum, $L_z$. One also observes that in these extremes points of $J_{norm}$ or, let us say, critical values of $b$, the $J_{norm}$ curve is a continuous one, whereas its derivative ($dJ/db$) is no longer well-defined. In addition, the singlet and triplet curves for $\sigma_x$ and $L_z$ present interchanged discontinuities in these critical values of the magnetic field. One can understand the continuous behaviour of $J_{norm}$ curve, as well as the discontinuities in the $\sigma_x$ and $L_z$ ones, by following the evolution of the lowest energy eigenstates as a function of $b$, i.e., by  following the lowest energy curves, and identifying the corresponding value of the $z$-component of the orbital angular momentum, $L_z$. This is what we show below. From now on we shall consider only the $quasi$-2D case which has more critical points for the same interval of $b \in [0, 7]$.

\begin{figure}[t]
%\vspace{-0,5 cm}
%\includegraphics[scale=0.23]{ES2D-todos-AAA.pdf}
\includegraphics[height=8 cm,width=\columnwidth]{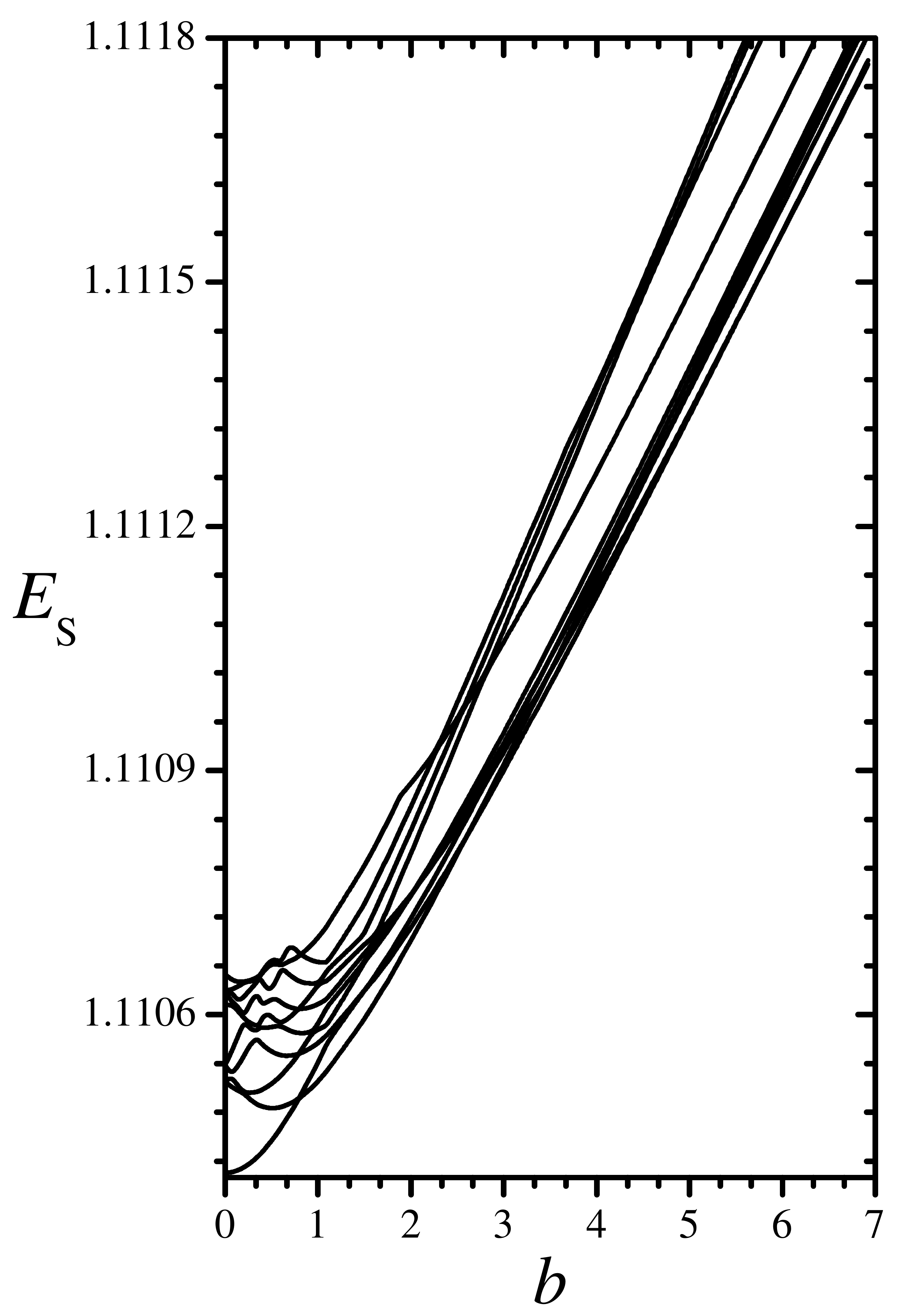}
\vspace{-0,5 cm}
\caption {It is plotted the first ten singlet energy curves as function of $b$ in the $quasi$-2D case of the two-electron QD. }
\label{ES2D-todos-AAA}
\end{figure}

\begin{figure}[t]
%\vspace{-0,5 cm}
%\includegraphics[scale=0.23]{ES2D-03-AAA.pdf}
\includegraphics[height=8 cm,width=\columnwidth]{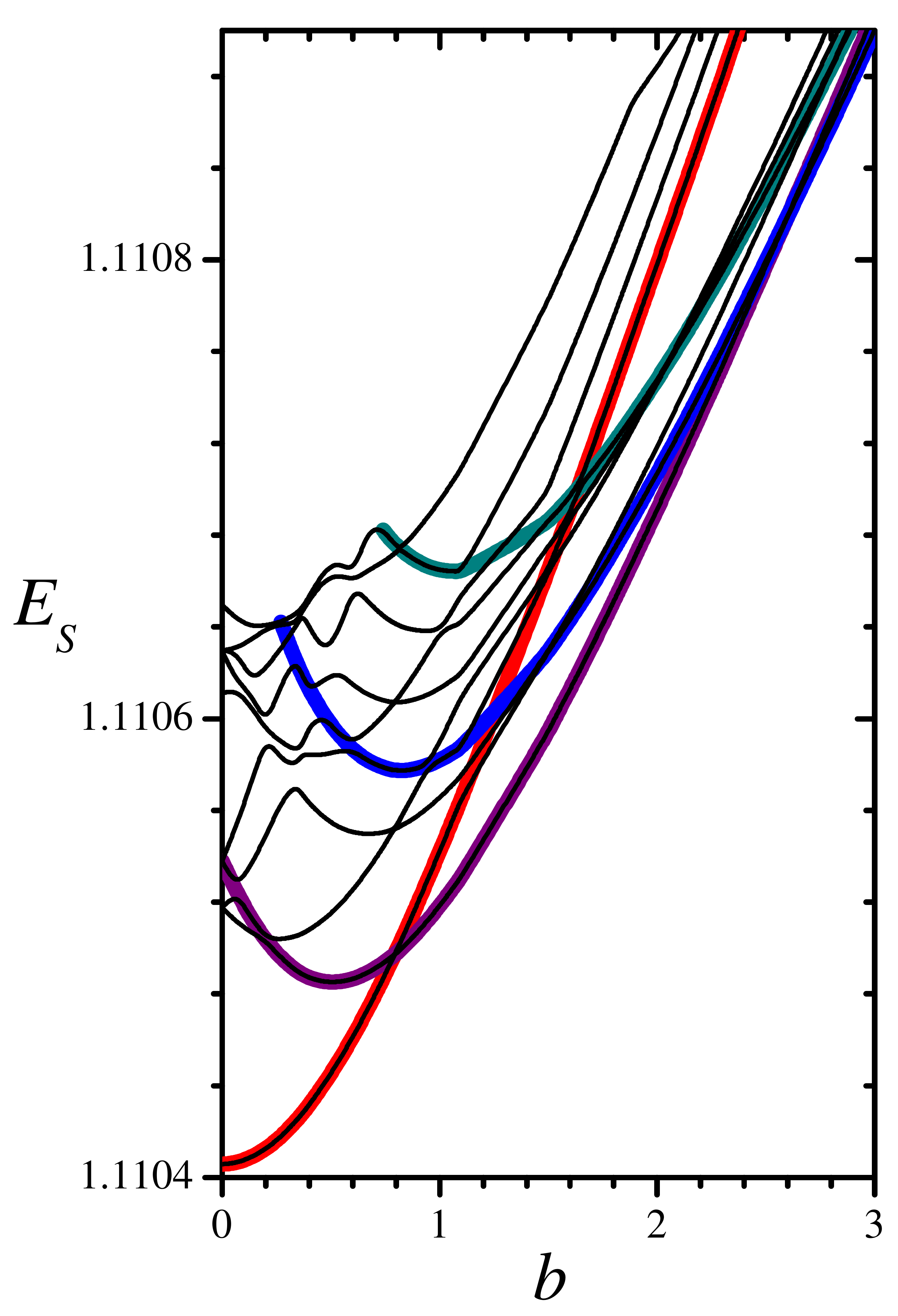}
\vspace{-0.5 cm}
\caption {A close-up of the previous figure. It is highlighted with thick lines (also colored in online version) the curves of the four eigenstates which play the role of the lowest energy singlet in the range $b \in [0,3]$.}
\label{ES2D-03-AAA}
\end{figure}

\begin{figure}[h]
%\vspace{-0,5 cm}
%\includegraphics[scale=0.23]{EFSLZ-2D-maior.pdf}
%\includegraphics[height=9.6 cm,width=\columnwidth]{EFSLZ-2D-maior.pdf} %=scale=0.23
\includegraphics[height=8 cm,width=\columnwidth]{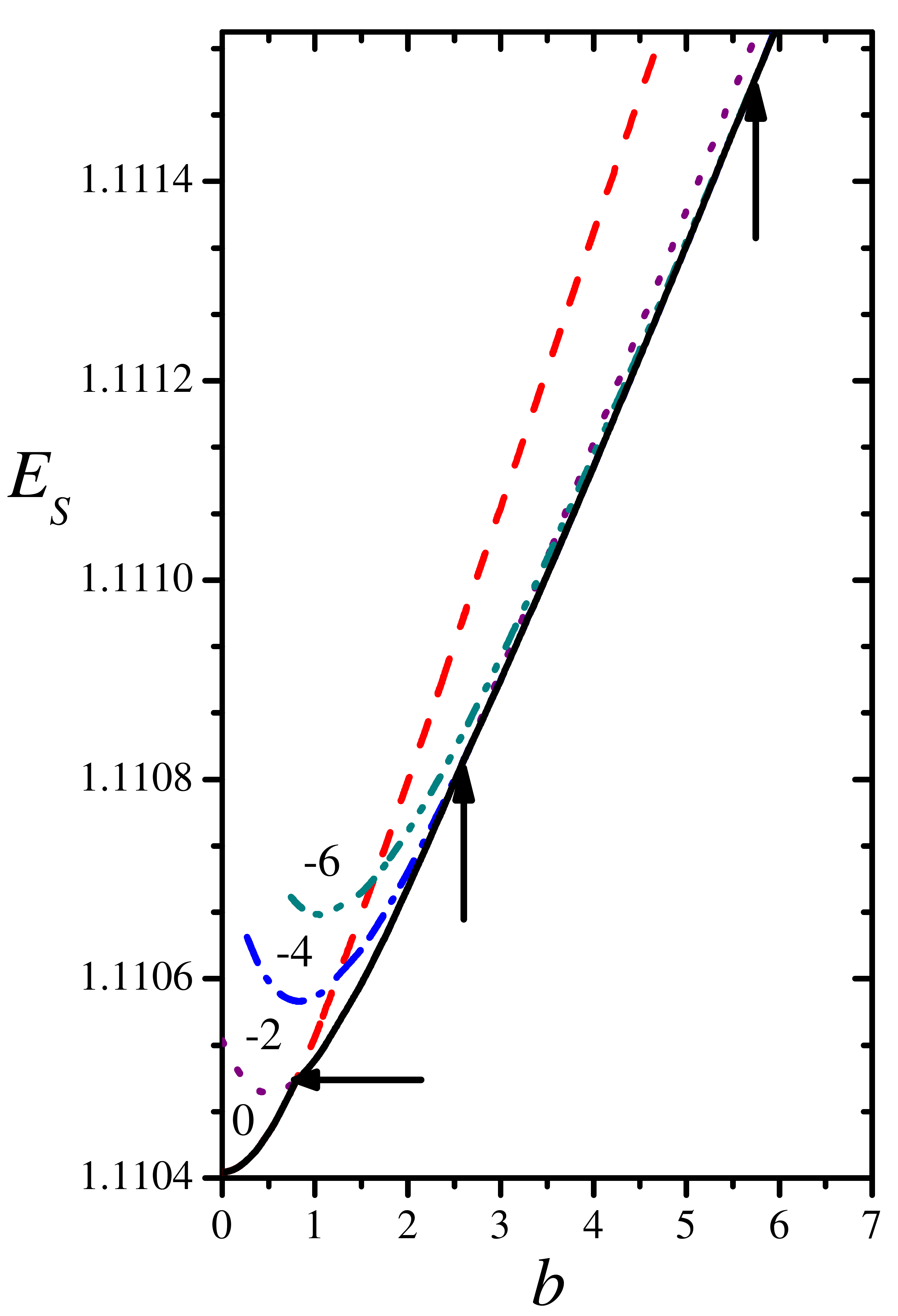}
\vspace{-0.5 cm}
\caption {The energy curves of the four eigenstates characterized by $L_z=0, -2, -4$ and $-6$ in the range of $b \in [0,7]$. Their curves correspond to the dashed (red), dotted (purple), dash-dotted (blue) and dash-dot-dotted (green) lines, respectively. The locations where these states change their role as the lowest energy singlet state (solid black line) are indicated by arrows  (colors in the online version). }
\label{EFSLz-2D}
\end{figure}

 Although an eigenstate of $\hat H$ is also an eigenstate of $\hat L_z$, when we order these states according to their energies for a specific value of $b$, it does not mean that the order of the lowest energy states remains unchanged as $b$ changes. In fact, what we observe, which is shown in Fig.~\ref{ES2D-todos-AAA} for the first ten singlet states ordered by energy, is a complete changing of the position of the eigenstates in respect to the energy. In Fig.~\ref  {ES2D-03-AAA} we display a close-up of the previous figure in order to make it easier the visualization.  The behaviour of the energy curves (solid black thin lines) may lead to a misunderstand of what is really happening when $b$ varies. One may believe that each of these lines describes the energy evolution of a specific eigenstate, but is wrong. The curves shown are just the energy curves of the first ten singlet states, computed point by point of $b$. To track the evolution of the energy of a specific eigenstate we must follow not only its energy but also its $L_z$; this is shown by the four solid (colored in  online version) thick lines. If one looks carefully, one can see that the top two thick curves each overlap with more than one thin energy curve. According to these lines, we observe that the role of the lowest energy singlet state is played by four different eigenstates with $L_z=0,-2,-4 {\rm \; and \;} -6$ as $b$ varies from 0 to 7. This is shown clearly in Fig.~\ref{EFSLz-2D}, where we plot only the curves corresponding to them and indicate with arrows the locations of changing from one eigenstate to other.

\begin{figure}[t]
%\vspace{-0,5 cm}
%\includegraphics[scale=0.23]{Fig6-Triplets2D_detalhes.pdf}
\includegraphics[height=8 cm,width=\columnwidth]{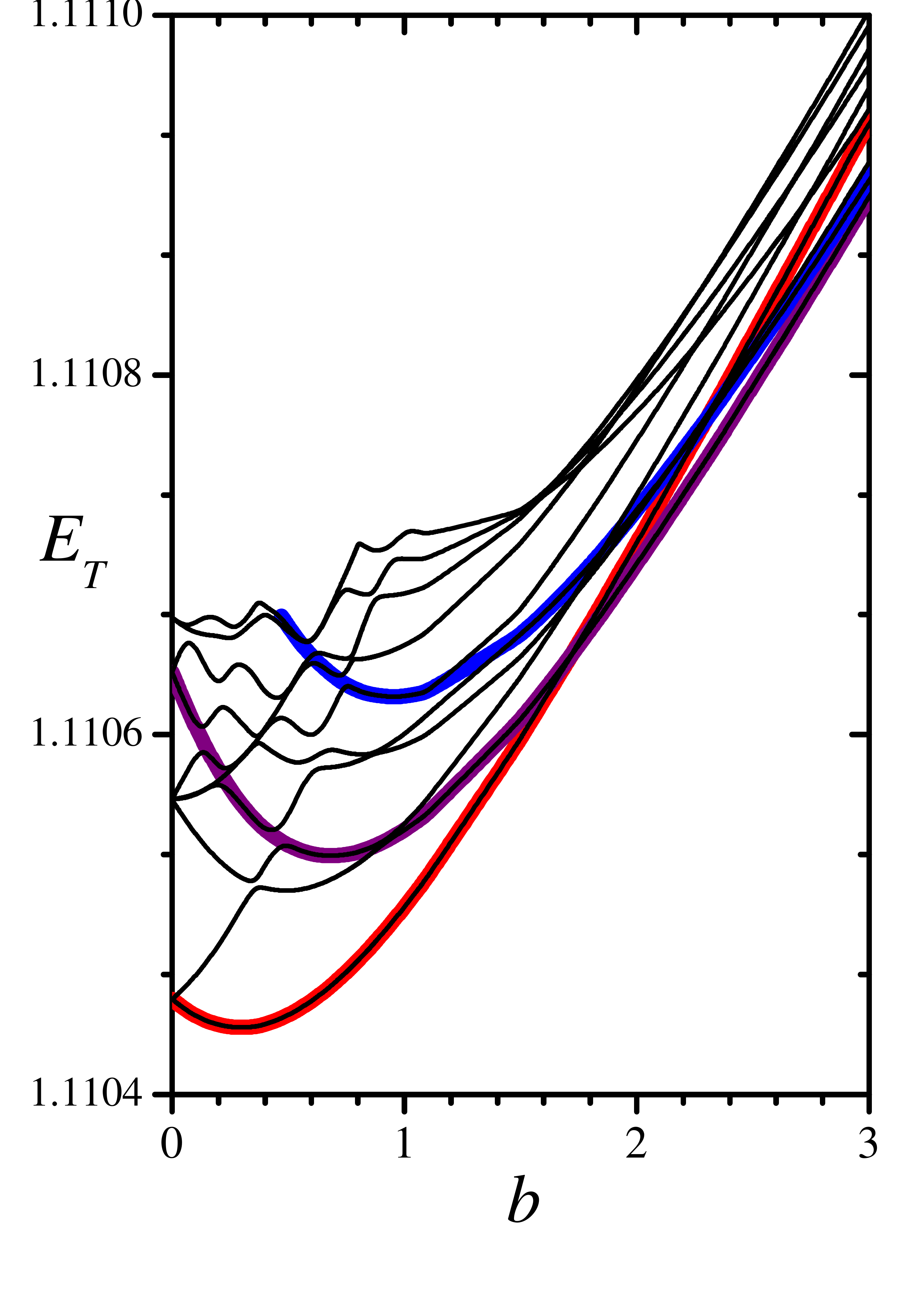}
\vspace{-0.5 cm}
\caption {Close-up of the first ten triplet energy curves as function of $b$. The energy curves of the eigenstates of $L_z=-1, -3$ and $-5$ are highlighted with thicker (colored in the online version) lines.}
\label{ET2D-03-AAA}
\end{figure}

\begin{figure}[h]
%\vspace{-0,5 cm}
%\includegraphics[scale=0.23]{Fig7-Triplets2D_Lz.pdf}
\includegraphics[height=8 cm,width=\columnwidth]{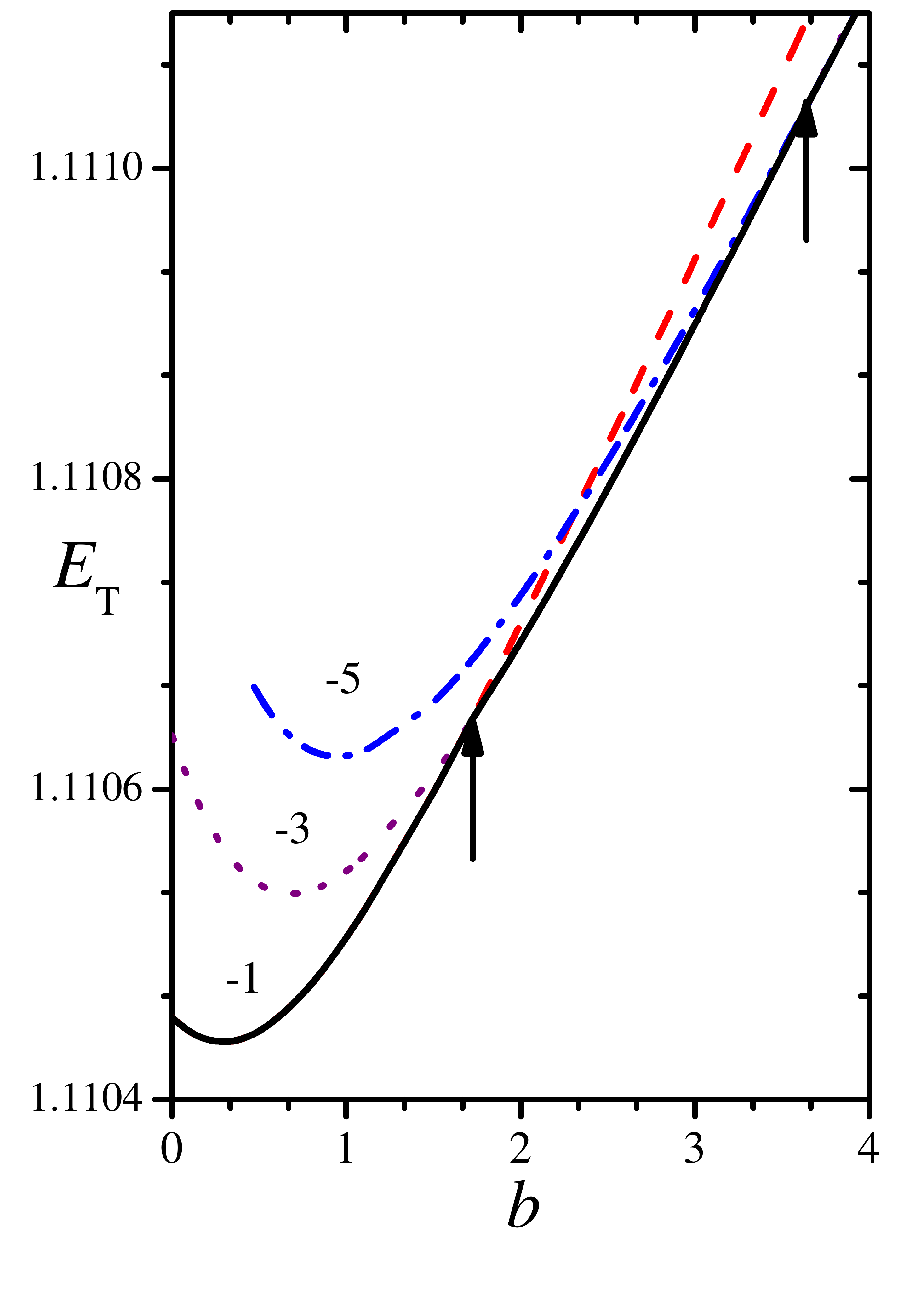}
\vspace{-0.5 cm}
\caption {From bottom to top it is shown the energy curves of the eigenstates with $L_z=-1, -3$ and $-5$, which correspond to the dashed (red), dotted (purple) and dash-dotted (blue) lines, respectively . The solid (black) line corresponds to the lowest energy triplet state (colors in the online version). It is also indicated with arrows the locations where occurs  the changes $L_z=-1 \rightarrow -3$ and $L_z=-3 \rightarrow -5$ as the lowest energy triplet state.}
\label{EFTLZ-2D}
\end{figure}

In Figs.~\ref{ET2D-03-AAA} and \ref{EFTLZ-2D} we display the corresponding behaviour of the triplet states. Fig.~\ref{ET2D-03-AAA} shows a close-up of the energies curves for the interval $b \in [0, 3]$ and one can see an analogous and complex behaviour of the first ten energies levels as function of $b$ observed in the singlet case. As in that one, the $i^{th}$ solid black thin line shows the evolution, as function of $b$, of the $i^{th}$ energy level and corresponds to different eigenstates along the $b-$axis. In the range of $b \in [0, 7]$ only the triplet states associated to $L_z=-1,-3 {\rm \; and \;} -5$ assume, at different values of $b$, the role of the lowest energy triplet state. Consequently, there are only two  changes in the $L_z$ value in this interval corresponding to $L_z=-1 \rightarrow -3$ and $L_z=-3 \rightarrow -5$. In Fig.~\ref{EFTLZ-2D} it is indicated with arrows the place where occurs these changes.

In conclusion, in this Letter we give a basic explanation for the oscillating properties of a two-electron QD that we discussed in a previous work \cite{maniero2020a} and that we identified as a manifestation of the {\it de Haas-van Alphen} effect, originally observed in the framework of diamagnetism of metals  \cite{deHaas1930}. This behaviour is a consequence of the mutual action of the Coulomb interaction and magnetic interaction on the two electrons in the QD, which lead to a complex evolution of the system eigenstates, as a function of the magnetic field $b$, making different eigenstates to assume in a certain interval of $b$ the condition of the lowest energy singlet and triplet states.

 %\pagebreak

%\begin{acknowledgments}
\ack
This work was partially supported by the Brazilian agencies CNPq,  CAPES, FAPESB and FAPERJ.
%\end{acknowledgments}

\vspace{1cm}
\bibliographystyle{unsrt}
\bibliography{RefQDots}

\end{document}